\documentclass[aps,prd,twocolumn,showpacs,superscriptaddress,groupedaddress]{revtex4}

\usepackage{graphicx}  
\usepackage{dcolumn}   
\usepackage{bm}        
\usepackage{amssymb}   
\usepackage{epsfig}
\usepackage{psfrag}
\usepackage{xspace}
\usepackage{multirow}
\usepackage{lscape}
\usepackage{comment}

\newcommand{\sigmatt}{\mbox{${\sigma}_{t\bar{t}}$}\xspace}

\newcommand{\dzero}     {D0}

\newcommand{\met}       {\mbox{\ensuremath{\slash\kern-.7emE_{T}}}}

\newcommand{\ttbar}     {\mbox{$t\bar{t}$}}

\newcommand{\ppbar}     {\mbox{$p\bar{p}$}}

\newcommand{\pythia}    {\sc{pythia}}
\newcommand{\alpgen}    {\sc{alpgen}}

\newcommand{\msbar}     {\mbox{$\overline{\rm MS}$}}
\newcommand{\mt}        {\mbox{\ensuremath{m_t}}}
\newcommand{\mtmsbar}   {\mbox{\ensuremath{m_t^{\overline{\rm MS}}}}}
\newcommand{\mtpole}	{\mbox{\ensuremath{m_t^{\rm pole}}}}
\newcommand{\mtmc}    {\mbox{\ensuremath{m_t^{\rm MC}}}}

\newcommand{\lumi}      {5.3~fb$^{-1}$}

\newcommand{\mnlo}     {164.8}
\newcommand{\ermnlo}   {^{+5.7}_{-5.4}}
\newcommand{\mta}      {163.0}
\newcommand{\ermta}    {^{+5.1}_{-4.6}}
\newcommand{\mtm}      {167.5}
\newcommand{\ermtm}    {^{+5.2}_{-4.7}}
\newcommand{\mtc}      {166.5}
\newcommand{\ermtc}    {^{+5.5}_{-4.8}}

\newcommand{\mtk}      {166.7}
\newcommand{\ermtk}    {^{+5.2}_{-4.5}}

\newcommand{\msbarpm}      {160.0}
\newcommand{\ermsbarpm}    {^{+4.8}_{-4.3}}
\newcommand{\msbarpa}      {154.5}
\newcommand{\ermsbarpa}    {^{+5.0}_{-4.3}}

\begin{document}

\hspace{5.2in} \mbox{Fermilab-Pub-11/177-E}

\title{Determination of the pole and $\overline{\bf MS}$ masses
of the top quark from the {\boldmath $t\bar{t}$} cross section}
\affiliation{Universidad de Buenos Aires, Buenos Aires, Argentina}
\affiliation{LAFEX, Centro Brasileiro de Pesquisas F{\'\i}sicas, Rio de Janeiro, Brazil}
\affiliation{Universidade do Estado do Rio de Janeiro, Rio de Janeiro, Brazil}
\affiliation{Universidade Federal do ABC, Santo Andr\'e, Brazil}
\affiliation{Instituto de F\'{\i}sica Te\'orica, Universidade Estadual Paulista, S\~ao Paulo, Brazil}
\affiliation{Simon Fraser University, Vancouver, British Columbia, and York University, Toronto, Ontario, Canada}
\affiliation{University of Science and Technology of China, Hefei, People's Republic of China}
\affiliation{Universidad de los Andes, Bogot\'{a}, Colombia}
\affiliation{Charles University, Faculty of Mathematics and Physics, Center for Particle Physics, Prague, Czech Republic}
\affiliation{Czech Technical University in Prague, Prague, Czech Republic}
\affiliation{Center for Particle Physics, Institute of Physics, Academy of Sciences of the Czech Republic, Prague, Czech Republic}
\affiliation{Universidad San Francisco de Quito, Quito, Ecuador}
\affiliation{LPC, Universit\'e Blaise Pascal, CNRS/IN2P3, Clermont, France}
\affiliation{LPSC, Universit\'e Joseph Fourier Grenoble 1, CNRS/IN2P3, Institut National Polytechnique de Grenoble, Grenoble, France}
\affiliation{CPPM, Aix-Marseille Universit\'e, CNRS/IN2P3, Marseille, France}
\affiliation{LAL, Universit\'e Paris-Sud, CNRS/IN2P3, Orsay, France}
\affiliation{LPNHE, Universit\'es Paris VI and VII, CNRS/IN2P3, Paris, France}
\affiliation{CEA, Irfu, SPP, Saclay, France}
\affiliation{IPHC, Universit\'e de Strasbourg, CNRS/IN2P3, Strasbourg, France}
\affiliation{IPNL, Universit\'e Lyon 1, CNRS/IN2P3, Villeurbanne, France and Universit\'e de Lyon, Lyon, France}
\affiliation{III. Physikalisches Institut A, RWTH Aachen University, Aachen, Germany}
\affiliation{Physikalisches Institut, Universit{\"a}t Freiburg, Freiburg, Germany}
\affiliation{II. Physikalisches Institut, Georg-August-Universit{\"a}t G\"ottingen, G\"ottingen, Germany}
\affiliation{Institut f{\"u}r Physik, Universit{\"a}t Mainz, Mainz, Germany}
\affiliation{Ludwig-Maximilians-Universit{\"a}t M{\"u}nchen, M{\"u}nchen, Germany}
\affiliation{Fachbereich Physik, Bergische Universit{\"a}t Wuppertal, Wuppertal, Germany}
\affiliation{Panjab University, Chandigarh, India}
\affiliation{Delhi University, Delhi, India}
\affiliation{Tata Institute of Fundamental Research, Mumbai, India}
\affiliation{University College Dublin, Dublin, Ireland}
\affiliation{Korea Detector Laboratory, Korea University, Seoul, Korea}
\affiliation{CINVESTAV, Mexico City, Mexico}
\affiliation{FOM-Institute NIKHEF and University of Amsterdam/NIKHEF, Amsterdam, The Netherlands}
\affiliation{Radboud University Nijmegen/NIKHEF, Nijmegen, The Netherlands}
\affiliation{Joint Institute for Nuclear Research, Dubna, Russia}
\affiliation{Institute for Theoretical and Experimental Physics, Moscow, Russia}
\affiliation{Moscow State University, Moscow, Russia}
\affiliation{Institute for High Energy Physics, Protvino, Russia}
\affiliation{Petersburg Nuclear Physics Institute, St. Petersburg, Russia}
\affiliation{Instituci\'{o} Catalana de Recerca i Estudis Avan\c{c}ats (ICREA) and Institut de F\'{i}sica d'Altes Energies (IFAE), Barcelona, Spain}
\affiliation{Stockholm University, Stockholm and Uppsala University, Uppsala, Sweden}
\affiliation{Lancaster University, Lancaster LA1 4YB, United Kingdom}
\affiliation{Imperial College London, London SW7 2AZ, United Kingdom}
\affiliation{The University of Manchester, Manchester M13 9PL, United Kingdom}
\affiliation{University of Arizona, Tucson, Arizona 85721, USA}
\affiliation{University of California Riverside, Riverside, California 92521, USA}
\affiliation{Florida State University, Tallahassee, Florida 32306, USA}
\affiliation{Fermi National Accelerator Laboratory, Batavia, Illinois 60510, USA}
\affiliation{University of Illinois at Chicago, Chicago, Illinois 60607, USA}
\affiliation{Northern Illinois University, DeKalb, Illinois 60115, USA}
\affiliation{Northwestern University, Evanston, Illinois 60208, USA}
\affiliation{Indiana University, Bloomington, Indiana 47405, USA}
\affiliation{Purdue University Calumet, Hammond, Indiana 46323, USA}
\affiliation{University of Notre Dame, Notre Dame, Indiana 46556, USA}
\affiliation{Iowa State University, Ames, Iowa 50011, USA}
\affiliation{University of Kansas, Lawrence, Kansas 66045, USA}
\affiliation{Kansas State University, Manhattan, Kansas 66506, USA}
\affiliation{Louisiana Tech University, Ruston, Louisiana 71272, USA}
\affiliation{Boston University, Boston, Massachusetts 02215, USA}
\affiliation{Northeastern University, Boston, Massachusetts 02115, USA}
\affiliation{University of Michigan, Ann Arbor, Michigan 48109, USA}
\affiliation{Michigan State University, East Lansing, Michigan 48824, USA}
\affiliation{University of Mississippi, University, Mississippi 38677, USA}
\affiliation{University of Nebraska, Lincoln, Nebraska 68588, USA}
\affiliation{Rutgers University, Piscataway, New Jersey 08855, USA}
\affiliation{Princeton University, Princeton, New Jersey 08544, USA}
\affiliation{State University of New York, Buffalo, New York 14260, USA}
\affiliation{Columbia University, New York, New York 10027, USA}
\affiliation{University of Rochester, Rochester, New York 14627, USA}
\affiliation{State University of New York, Stony Brook, New York 11794, USA}
\affiliation{Brookhaven National Laboratory, Upton, New York 11973, USA}
\affiliation{Langston University, Langston, Oklahoma 73050, USA}
\affiliation{University of Oklahoma, Norman, Oklahoma 73019, USA}
\affiliation{Oklahoma State University, Stillwater, Oklahoma 74078, USA}
\affiliation{Brown University, Providence, Rhode Island 02912, USA}
\affiliation{University of Texas, Arlington, Texas 76019, USA}
\affiliation{Southern Methodist University, Dallas, Texas 75275, USA}
\affiliation{Rice University, Houston, Texas 77005, USA}
\affiliation{University of Virginia, Charlottesville, Virginia 22901, USA}
\affiliation{University of Washington, Seattle, Washington 98195, USA}
\author{V.M.~Abazov} \affiliation{Joint Institute for Nuclear Research, Dubna, Russia}
\author{B.~Abbott} \affiliation{University of Oklahoma, Norman, Oklahoma 73019, USA}
\author{B.S.~Acharya} \affiliation{Tata Institute of Fundamental Research, Mumbai, India}
\author{M.~Adams} \affiliation{University of Illinois at Chicago, Chicago, Illinois 60607, USA}
\author{T.~Adams} \affiliation{Florida State University, Tallahassee, Florida 32306, USA}
\author{G.D.~Alexeev} \affiliation{Joint Institute for Nuclear Research, Dubna, Russia}
\author{G.~Alkhazov} \affiliation{Petersburg Nuclear Physics Institute, St. Petersburg, Russia}
\author{A.~Alton$^{a}$} \affiliation{University of Michigan, Ann Arbor, Michigan 48109, USA}
\author{G.~Alverson} \affiliation{Northeastern University, Boston, Massachusetts 02115, USA}
\author{G.A.~Alves} \affiliation{LAFEX, Centro Brasileiro de Pesquisas F{\'\i}sicas, Rio de Janeiro, Brazil}
% ...
%
%
\collaboration{The D0 Collaboration\footnote{with visitors from
$^{a}$Augustana College, Sioux Falls, SD, USA,
$^{b}$The University of Liverpool, Liverpool, UK,
$^{c}$SLAC, Menlo Park, CA, USA,
$^{d}$University College London, London, UK,
$^{e}$Centro de Investigacion en Computacion - IPN, Mexico City, Mexico,
$^{f}$ECFM, Universidad Autonoma de Sinaloa, Culiac\'an, Mexico,
and 
$^{g}$Universit{\"a}t Bern, Bern, Switzerland.
}} \noaffiliation
\vskip 0.25cm

\date{April 14, 2011}

\begin{abstract}

We use higher-order quantum chromodynamics calculations
to extract the mass of the top quark from the \ttbar\ cross section
measured in the lepton+jets channel in $\ppbar$ collisions at
$\sqrt{s}=1.96$~TeV  
using \lumi\ of integrated luminosity collected by the \dzero\ experiment at the
Fermilab Tevatron Collider. 
The extracted top quark pole mass and \msbar\ mass are
compared to the current Tevatron average top quark mass obtained from
direct measurements.

\end{abstract}


\maketitle

\newpage

The mass of the top quark (\mt) has been measured with a precision of
0.6\%, and its 
current Tevatron average value is $m_t =173.3 \pm
1.1$~GeV~\cite{tevatron_wa}. Beyond leading-order quantum
chromodynamics (LO QCD), the mass of the top quark is a 
convention-dependent parameter. Therefore, it is important to know how
to interpret 
this experimental result in terms of
renormalization conventions~\cite{hoang} if the value is to be used
as an input to higher-order QCD calculations or in fits of electroweak
precision observables and the resulting indirect Higgs boson mass
bounds~\cite{ew_fits}. 
The definition of mass in field theory can be divided into two
categories~\cite{mcpaper_formassdef}: 
({\it i}) driven by long-distance behavior, which corresponds to the
pole-mass scheme, 
and ({\it ii}) driven by short-distance behavior, which,
for example, is represented by the \msbar\ mass scheme. The difference
between the masses in 
different schemes can be calculated as a perturbative series in
$\alpha_s$. 
However, the concept of the
pole mass is ill-defined, since there is no pole in the quark
propagator in a confining theory such as QCD~\cite{willenbrock}.

There are two approaches to directly measure \mt\
from the 
reconstruction of the final states in decays of top-antitop ($t\bar{t}$)
pairs. One is based on a comparison of Monte Carlo (MC) templates for
different assumed values of \mt\ with distributions
of kinematic quantities measured in data. In the second approach,
\mt\ is extracted from the reconstruction
of the final states in data using a calibration curve obtained
from MC simulation. In both cases the quantity measured in
data therefore corresponds to the top quark mass scheme used in
the MC simulation, which we refer to as \mtmc. 

Current
MC simulations
are performed in LO QCD, and higher order effects are simulated
through parton showers at modified leading logarithms (LL) level.
In principle, it is not possible to establish
a direct connection between \mtmc\ and any other mass scheme, such as
the pole or \msbar\ mass scheme, without calculating the parton showers
to at least next-to-leading logarithms (NLL)
accuracy. However, it has been argued that \mtmc\
should be close to the 
pole mass~\cite{jetmass_scheme,jetmass_footnote}.
The relation between \mtmc\ and
the top quark pole mass (\mtpole) or \msbar\ mass (\mtmsbar) is still
under theoretical investigation.
In calculations such as in Ref.~\cite{ew_fits} it is assumed that \mtmc\
measured at the Tevatron is equal to \mtpole.

In this Letter, we extract the pole mass at the scale of the pole mass,
\mtpole(\mtpole), and the \msbar\ mass at the scale of the \msbar\
mass, \mtmsbar(\mtmsbar), 
comparing the measured inclusive \ttbar\ production cross section
\sigmatt\ with fully
inclusive calculations at higher-order QCD that involve an unambiguous
definition of \mt\ and compare our results to
\mtmc.  
This extraction provides an important test of the mass
scheme as applied in MC simulations and gives complementary
information, with different sensitivity to theoretical and
experimental uncertainties than the 
direct measurements of \mtmc\ that rely on kinematic details of the mass
reconstruction.

We use the measurement of \sigmatt\ in the lepton+jets channel in $\ppbar$ collisions at
$\sqrt{s}=1.96$~TeV using \lumi\ of integrated luminosity collected
by the \dzero\ experiment~\cite{xsec_paper}.
We calculate likelihoods for \sigmatt\ as a function of \mt,
and use higher-order QCD predictions based on
the pole-mass or the \msbar-mass conventions
to extract \mtpole or \mtmsbar, respectively.

The criteria applied to select the sample of \ttbar\
candidates used in the cross section measurement introduce
a dependence of the signal acceptance, and therefore of the
measured value of \sigmatt, on the assumed value
of \mtmc. This dependence is studied
using MC samples of \ttbar\ events generated at different
values of \mtmc\ in intervals of at least 5~GeV and
is found to be much weaker than the dependence of
the theoretical calculation of \sigmatt\ on \mt.
The $t\bar{t}$ signal is simulated with the {\alpgen} event 
generator~\cite{Alpgen}, and parton evolution is 
simulated with {\pythia}~\cite{Pythia}. Jet-parton matching is
applied to avoid double-counting of partonic event
configurations~\cite{matching}. 
The resulting measurement of \sigmatt\ can be described by 
\begin{eqnarray} \label{eq_massfit}
\sigma_{t\bar{t}}(\mtmc) &=& 
\frac{1}{(\mtmc)^4} [ a + b\ (\mtmc -m_0)  \\ \nonumber
&+& c\ (\mtmc -m_0)^2 + d\  (\mtmc -m_0)^3]\, ,
\end{eqnarray} 
where $\sigma_{t\bar{t}}$ and \mtmc\ are in~pb and~GeV, respectively,
$m_0 = 170$~GeV, and $a$, $b$, $c$, $d$ are free parameters. For the
mass extraction, we consider the 
experimental \ttbar\ cross section measured using the $b$-jet identification 
technique~\cite{xsec_paper}. 
This \sigmatt\ determination provides the weakest 
dependence on \mtmc\ of the results presented in
Ref.~\cite{xsec_paper}, which leads to a smaller uncertainty on the
extracted $m_{t}$,  
and thereby reduces the ambiguity of whichever convention 
(here pole or \msbar) best reflects \mtmc.
When using $b$-tagging, the data sample is split into events with $0,
1$ or $> 1$ 
$b$-tagged jets, and the numbers  
of events in each of the three categories, corrected for mass-dependent
acceptance, yield
the measurement of \sigmatt.
The other methods used in Ref.~\cite{xsec_paper} 
rely on additional topological information that introduces
a stronger dependence of the measured \sigmatt\ on \mtmc. They are
therefore not used in this analysis.
The parameters derived from a fit of \sigmatt\ to Eq.~(\ref{eq_massfit})
are: $a=6.95 \times 10^9 \ {\rm pb}\, {\rm GeV}^4$, $b=1.25 \times
10^8 \ {\rm pb}\, {\rm GeV}^3$, $c=1.16 \times 10^6 \ {\rm pb}\,
{\rm GeV}^2$, and $d=-2.55 \times 10^3 \ {\rm pb}\, {\rm GeV}$.
Possible fit shape changes due to the uncertainties on these parameters
are small compared to the experimental uncertainties
on the \sigmatt\ measurement which are almost fully correlated
between different \mt.
For \mtmc=172.5~GeV, we measure $\sigmatt = 8.13 ^{+1.02} _{-0.90} $~pb~\cite{xsec_paper}.

We compare the obtained parameterization to a pure next-to-leading-order (NLO)
QCD~\cite{SMtheory_N} calculation,
to a calculation including NLO QCD
and all higher-order soft-gluon resummations in NLL~\cite{SMtheory_C},
to a calculation including also 
all higher-order soft-gluon 
resummations in next-to-next-to-leading logarithms (NNLL)~\cite{SMtheory_A}
and to two approximations of the
next-to-next-to-leading-order (NNLO) QCD cross section that include
next-to-next-to-leading logarithms (NNLL) relevant in NNLO
QCD~\cite{SMtheory_M,SMtheory_K}. 
The computations in Ref.~\cite{SMtheory_M} were obtained using the
program documented in Ref.~\cite{hathor}.

Following the method of Refs.~\cite{ljets,dilepton}, we extract the
most probable \mt\ values and their 68\% C.L. bands for the
pole-mass and \msbar-mass conventions by computing the most probable value of a normalized joint-likelihood function:
\begin{equation}
L(\mt) = \int f_{\mathrm{exp}} (\sigma | \mt) \, \left[ f_{\rm scale}
(\sigma | \mt) \otimes f_{\rm PDF} (\sigma | \mt) \right] \, d\sigma. 
\label{eq:fmt}
\end{equation}
The first term $f_{\mathrm{exp}}$ corresponds to 
a function for the measurement constructed from a Gaussian function with
mean value given by Eq.~(\ref{eq_massfit})
and with standard deviation (sd) equal to the total experimental uncertainty 
which is described in detail in Ref.~\cite{xsec_paper}.
The second term $f_{\mathrm{scale}}$  in Eq.~(\ref{eq:fmt}) is a theoretical
likelihood formed from the uncertainties on the 
renormalization and factorization  
scales of QCD, which are taken to be equal,
and varied up and down by a factor of two from the default
value.
Within this range, $f_{\mathrm{scale}}$ is taken to be
constant~\cite{SMtheory_N,SMtheory_C,SMtheory_A,SMtheory_M,SMtheory_K}.
It is convoluted with a term that represents
the uncertainty of parton density functions (PDFs), taken to be
a Gaussian function, with rms equal to the uncertainty determined
in Refs.~\cite{SMtheory_N,SMtheory_C,SMtheory_A,SMtheory_M,SMtheory_K}. 
Table~\ref{tab:theory_uncert} summarizes the theoretical predictions from
different calculations 
for $\mtpole=175$~GeV used as an input to the likelihood fit.    
\begin{table}
\begin{center}
\caption{Theoretical predictions for \sigmatt\ with uncertainties
$\Delta \sigma$ due to scale dependence and PDFs at the Tevatron for
\mtpole=175~GeV from different theoretical
calculations used as input to the mass extraction. Note
that Refs.~\cite{SMtheory_N} and \cite{SMtheory_C}   
use the CTEQ6.6 PDF set~\cite{cteq} 
while Refs.~\cite{SMtheory_A}, \cite{SMtheory_M}, and \cite{SMtheory_K} use the
MSTW08 PDF set~\cite{mstw}.} 
\vspace{0.2cm}
\begin{tabular}{c|ccc} 
\hline \hline
Theoretical prediction & \sigmatt\ (pb) & $\Delta \sigma_{\rm scale}$ (pb) & $\Delta \sigma_{\rm PDF}$ (pb) \\ \hline \\[-10pt]
NLO \cite{SMtheory_N}	            & 6.39 & $^{+0.33}_{-0.70}$ & $^{+0.35}_{-0.35}$  \\[2pt] \\[-10pt]
NLO+NLL \cite{SMtheory_C}	    & 6.61 & $^{+0.26}_{-0.46}$ & $^{+0.44}_{-0.34}$  \\[2pt] \\[-10pt]
NLO+NNLL \cite{SMtheory_A}          & 5.93 & $^{+0.18}_{-0.17}$ & $^{+0.30}_{-0.22}$  \\[2pt] \\[-10pt]
Approximate NNLO \cite{SMtheory_M}  & 6.71 & $^{+0.28}_{-0.37}$ & $^{+0.17}_{-0.12}$  \\[2pt] \\[-10pt]
Approximate NNLO \cite{SMtheory_K}  & 6.66 & $^{+0.11}_{-0.06}$ &
$^{+0.42}_{-0.35}$  \\[2pt] \hline \hline
\end{tabular}
\label{tab:theory_uncert}
\end{center}
\end{table}

In Refs.~\cite{SMtheory_N,SMtheory_C,SMtheory_A,SMtheory_M,SMtheory_K}  
\sigmatt\ is calculated as a function of \mtpole\ and, consequently,
comparing the measured $\sigmatt(\mtmc)$ to these theoretical
predictions provides a 
value of \mtpole. 
Therefore, we extract \mtpole\ ({\it i}) assuming that the definition of \mtmc\
is equivalent to \mtpole, and ({\it ii})
taking \mtmc\ to be equal to \mtmsbar\ to estimate the 
effect of interpreting \mtmc\ as 
any other mass definition. 
%
%
\begin{figure}[ht]
\flushright
\hspace*{-1.5cm}\includegraphics[width=0.51\textwidth]{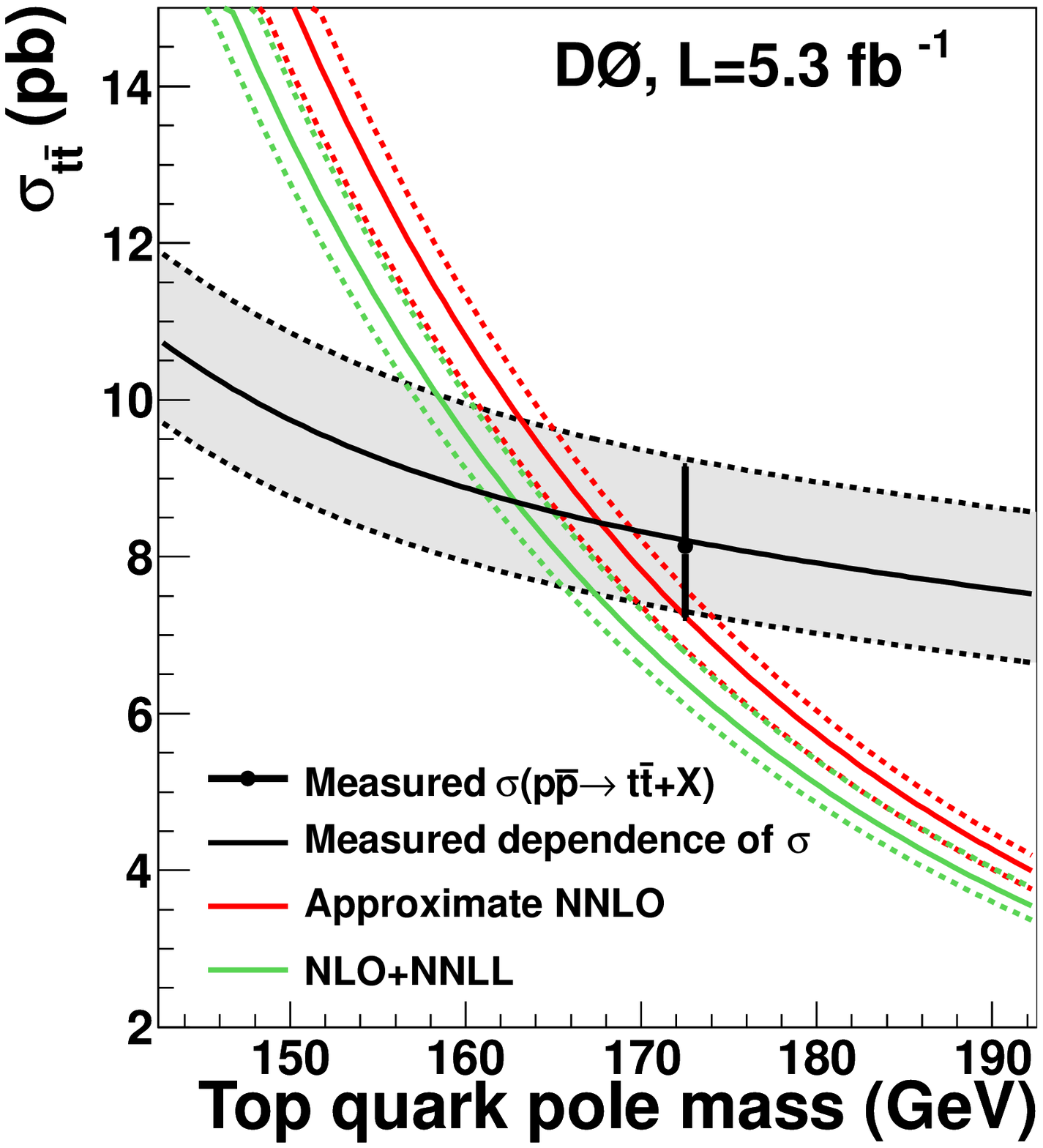}
\begin{center}
\caption{(Color online) Measured $\sigmatt$ and
theoretical NLO+NNLL~\cite{SMtheory_A} and approximate
NNLO~\cite{SMtheory_M} calculations of \sigmatt\ as 
a function of \mtpole, assuming that $\mtmc = \mtpole$. 
The colored dashed lines  
represent the uncertainties for the two theoretical calculations
from the choice of the PDF 
and the renormalization and factorization scales (added quadratically).
The theoretical calculation of Ref.~\cite{SMtheory_K} (not
displayed) agrees with Ref.~\cite{SMtheory_M} within 1\% both in mean value
and uncertainty.
The point shows the measured \sigmatt\ for $\mtmc=172.5$~GeV, the black curve is
the fit to Eq.~(\ref{eq_massfit}), and the gray  
band corresponds to the total experimental
uncertainty.\label{fig:pole_scenario1}} 
\end{center}
\end{figure}
%
%
\begin{table}[t]
\begin{center}
\caption{Values of \mtpole, with their 68\% C.L. uncertainties,
extracted for different 
predictions of \sigmatt. The results assume that $\mtmc = \mtpole$
(left column). The right column shows the 
change $\Delta m_{t}^{\rm pole}$ between these results if it is
assumed that $\mtmc = \mtmsbar$.
The combined experimental and theoretical uncertainties are shown.}
\vspace{0.2cm}
\begin{tabular}{c|cc} 
\hline \hline \\[-9pt] 
Theoretical prediction          & {$m_{t}^{\rm pole}$ (GeV)} &
{$\Delta m_{t}^{\rm pole}$ (GeV)} \\[2pt] \hline \\[-8pt]
MC mass assumption  & $m_{t}^{\rm MC} = m_{t}^{\rm pole}$ & $m_{t}^{\rm MC} = m_{t}^{\overline{\rm MS}}$ \\[2pt] \hline \\[-8pt]
NLO \cite{SMtheory_N}              &   $\mnlo \ermnlo$ & $-3.0$ \\[2pt]
NLO+NLL \cite{SMtheory_C}          & $\mtc \ermtc$ & $-2.7$ \\[2pt]
NLO+NNLL \cite{SMtheory_A}         & $\mta \ermta$ & $-3.3$ \\ [2pt]
Approximate NNLO \cite{SMtheory_M} & $\mtm \ermtm$ & $-2.7$ \\ [2pt]
Approximate NNLO \cite{SMtheory_K} & $\mtk \ermtk$ & $-2.8$ \\ [1pt] \hline \hline
\end{tabular}
\label{tab:pole}
\end{center}
\end{table}
%
%
%
%
For case ({\it i}), 
Fig.~\ref{fig:pole_scenario1} shows the parameterization  of the
measured  and the predicted
$\sigmatt(\mtpole)$~\cite{SMtheory_A,SMtheory_M,SMtheory_K}.  
The results for the determination of \mtpole\ are given
in Table~\ref{tab:pole}. 
In case ({\it ii}) the cross section predictions
use the pole-mass convention, 
and the value of $\mtmc=\mtmsbar$ is converted to \mtpole\ using the 
relationship at the three-loop level~\cite{willenbrock,3loop}:
\begin{eqnarray}
\label{eq:trans}
& &m_{t}^{\rm pole} = m_{t}^{\overline{\rm MS}} (m_{t}^{\overline{\rm MS}}) \, \Big[ 1 + \frac{4}{3}
\frac{\overline{\alpha}_s (m_{t}^{\overline{\rm MS}})}{\pi} \\ \nonumber
&+& (-1.0414 N_L + 13.4434) \left(
\frac{\overline{\alpha}_s(m_{t}^{\overline{\rm MS}})}{\pi} \right)^2
\\ \nonumber 
&+& (0.6527 N_L^2 - 26.655 N_L + 190.595) \left(
\frac{\overline{\alpha}_s(m_{t}^{\overline{\rm MS}})}{\pi} \right)^3 \, \Big] \,\, ,
\end{eqnarray}
where $\overline{\alpha}_s$ is the strong coupling in the \msbar\
scheme, and $N_L=5$ is the number of light quark flavors.
The strong coupling $\overline{\alpha}_s(m_{t}^{\rm pole})$
is taken at the three-loop level from Ref.~\cite{bethke}. 
By iteratively rederiving the \msbar\ mass using Eq.~(\ref{eq:trans})
$\overline{\alpha}_s(m_{t}^{\rm pole})$ 
is transformed into $\overline{\alpha}_s 
(m_{t}^{\overline{\rm MS}})$ leading to a difference of only 0.1~GeV
to the final extraction of \mtmsbar.
For $m_{t}^{\rm
  pole} = 173.3$~GeV, the \msbar\ mass $m_{t}^{\overline{\rm MS}}
  (m_{t}^{\overline{\rm MS}})$ is lower by 9.8~GeV.
With this change of the \mtmc\ interpretation in Eq.~(\ref{eq_massfit}), we
form a new likelihood $f_{\mathrm{exp}} (\sigma | \mt)$ and extract
\mtpole\ using Eq.~(\ref{eq:fmt}).
The difference $\Delta m_{t}^{\rm pole}$ between assuming $\mtmc = \mtpole$ and
$\mtmc = \mtmsbar$ is
given in Table~\ref{tab:pole}.
Given the uncertainties, 
interpreting \mtmc\ as either \mtpole\ or as \mtmsbar\ has
no significant bearing on the value of the extracted \mt. We include
half of this difference symmetrically in the systematic uncertainties. 
As a result we extract $\mtpole = \mta ^{+5.4}
_{-4.9}$~GeV using the NLO+NNLL calculation of Ref.~\cite{SMtheory_A} and $\mtpole = \mtm
^{+5.4}_{-4.9}$~GeV 
using the approximate NNLO calculation of Ref.~\cite{SMtheory_M}.
Our measurement of \mtpole\ based on the approximate NNLO cross
section calculation is consistent within 1~sd with the Tevatron
measurement of \mt\ 
from direct reconstruction of top quark decay products, $m_t =173.3
\pm 1.1$~GeV~\cite{tevatron_wa}. The result
based on the NLO+NNLL calculation is consistent within 2~sd.

Calculations of the \ttbar\ cross
section~\cite{SMtheory_A,SMtheory_M} have also been performed as a
function of \mtmsbar. Comparing the dependence of the
measured \sigmatt\ 
to theory as a function of \mt\ provides an estimate of \mtmsbar. 
We note that a previous extraction of \mtmsbar~\cite{SMtheory_M} 
ignored the \mt\ dependence of the measured \sigmatt. 

\begin{figure}[h]
\flushright
\hspace*{-1.5cm}\includegraphics[width=0.51\textwidth]{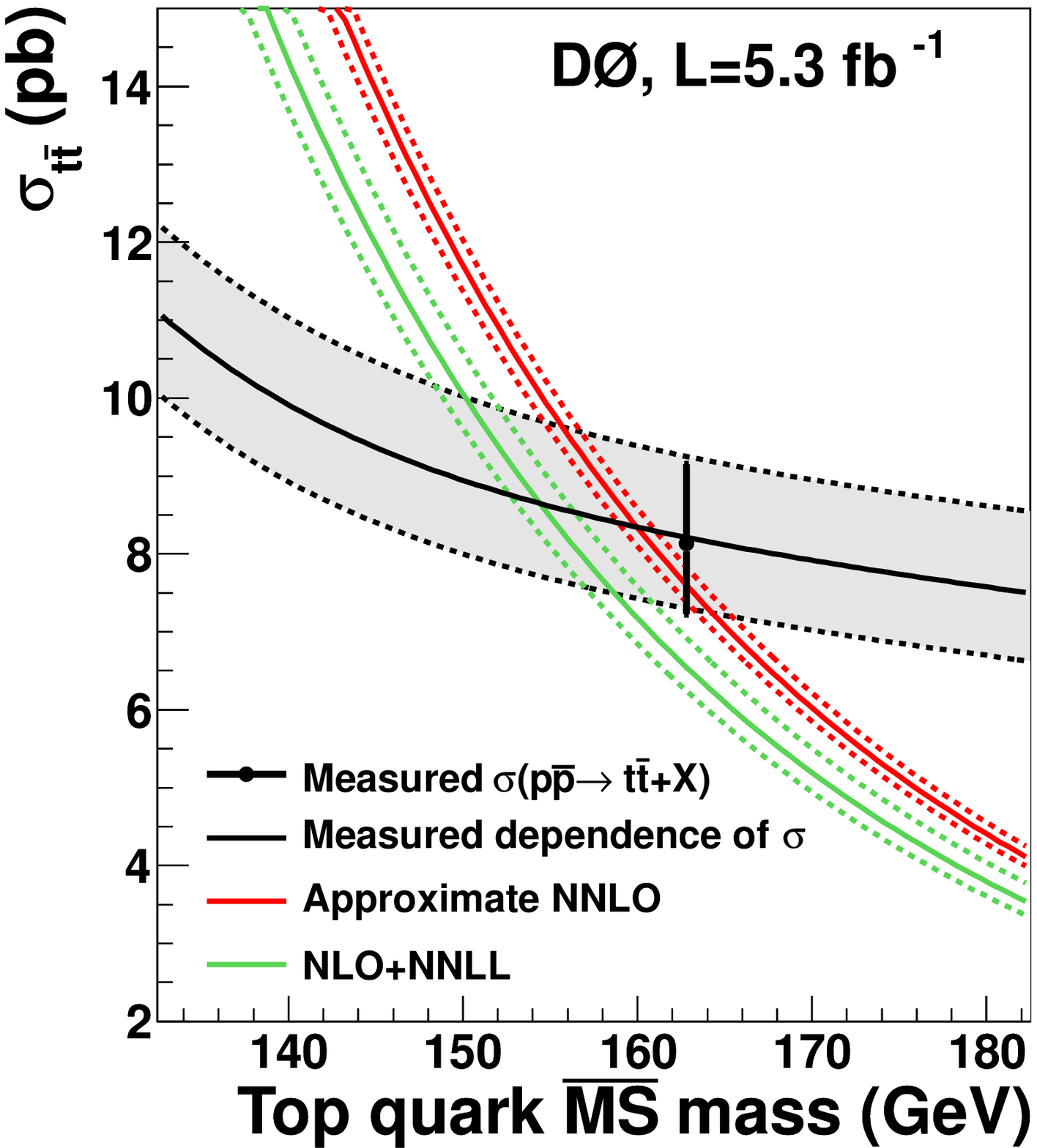}
\begin{center}
\caption{(Color online) Measured $\sigmatt$ and
theoretical NLO+NNLL~\cite{SMtheory_A} and approximate
NNLO~\cite{SMtheory_M} calculations of \sigmatt\ as 
a function of \mtmsbar, assuming that $\mtmc = \mtpole$.
The colored dashed lines  
represent the uncertainties for the two theoretical calculations from
the choice of the PDF 
and the renormalization and factorization scales (added quadratically).
The point shows the measured \sigmatt\ for \mtmc=172.5~GeV, the black curve is
the fit to Eq.~(\ref{eq_massfit}), and the gray  
band corresponds to the total experimental
uncertainty.\label{fig:msbar_scenario1}} 
\end{center}
\end{figure}

We extract the value of \mtmsbar, again, for two cases:
({\it i}) assuming that the definition of \mt\ implemented in the MC
simulation is  
equal to \mtpole, and ({\it ii}) assuming that \mtmc\
corresponds to \mtmsbar. 
For case ({\it i}), \mtpole\ must first be converted to \mtmsbar\
using Eq.~(\ref{eq:trans}). 
Figure~\ref{fig:msbar_scenario1} shows the measured \sigmatt\
as a function of \mtmsbar, together with 
the calculation that includes NLO+NNLL QCD
resummation~\cite{SMtheory_A} and the approximate NNLO calculation~\cite{SMtheory_M}.

The results for the extracted values of \mtmsbar\ are
given in Table~\ref{tab:msbar}.   
%
\begin{table}[ht!]
\begin{center}
\caption{Values of \mtmsbar, with their 68\% C.L. uncertainties,
extracted for different 
theoretical predictions of \sigmatt. The results assume that \mtmc\
corresponds to \mtpole\ (left column). The right column shows the
change $\Delta m_{t}^{\overline{\rm MS}}$ between these results if it
is assumed that $\mtmc = \mtmsbar$.
The combined experimental and theoretical uncertainties are shown.}
\vspace{0.2cm}
\begin{tabular}{c|cc} 
\hline \hline \\[-9pt] 
Theoretical prediction & {$m_{t}^{\overline{\rm MS}}$ (GeV)}  &
{$\Delta m_{t}^{\overline{\rm MS}}$ (GeV)} \\[2pt] \hline \\[-8pt]
MC mass assumption  & $m_{t}^{\rm MC} = m_{t}^{\rm pole}$ & $m_{t}^{\rm MC} = m_{t}^{\overline{\rm MS}}$ \\[2pt] \hline \\[-8pt]
NLO+NNLL~\cite{SMtheory_A} & $\msbarpa \ermsbarpa$ & $-2.9$ \\ [2pt]
Approximate NNLO~\cite{SMtheory_M} & $\msbarpm \ermsbarpm$ & $-2.6$ \\ [1pt] \hline \hline
\end{tabular}
\label{tab:msbar}
\end{center}
\end{table}
%
%
\begin{figure}[ht!]
\flushright
\hspace*{-1.5cm}\includegraphics[width=0.51\textwidth]{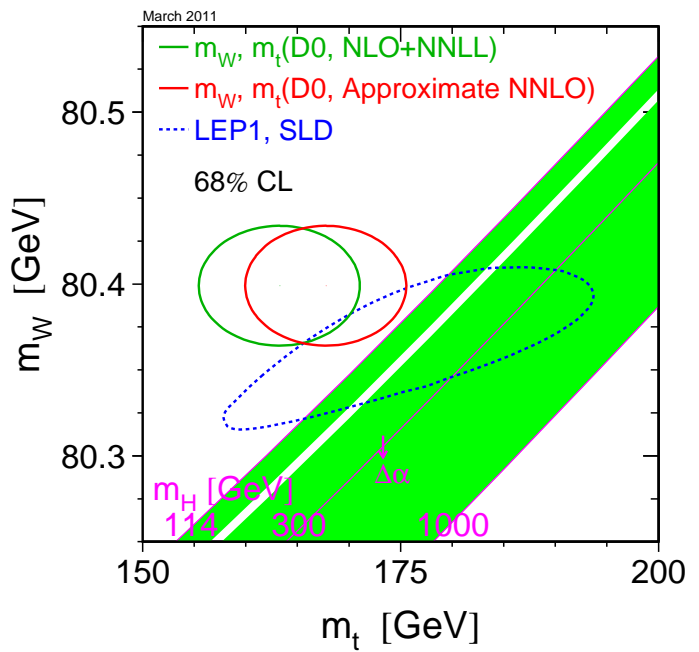}
\begin{center}
\caption{(Color online) Constraints on the $W$ boson mass from the LEP-II/Tevatron
experiments and the top quark pole mass extracted from the \ttbar\ cross section in
NLO+NNLL~\cite{SMtheory_A} (green contour) and approximate NNLO
~\cite{SMtheory_M} (red contour). This is compared to the indirect
constraints on the $W$ boson mass and 
the top quark mass based on LEP-I/SLD data (dashed contour).
In both cases the 68\% CL contours are given. Also shown is the SM
relationship for the masses as a function of the Higgs mass in the
region favoured by theory ($< 1000$~GeV) and not excluded by direct
searches (114~GeV to 158~GeV and $> 173$~GeV). The arrow labelled
$\Delta\alpha$ shows the variation of this relation if $\alpha(m_Z^2)$ is varied
between $-1$ and $+1$ sd. This variation gives an additional
uncertainty to the SM 
band shown in the figure.}
\label{fig:w11}
\end{center}
\end{figure}

In case ({\it ii}), we assume that the mass definition in
the MC simulation corresponds to the \msbar\ mass.
We set $\mtmc = \mtmsbar$ in Eq.~(\ref{eq:fmt}), form a new
likelihood $f_{\mathrm{exp}} (\sigma | \mt)$ and extract 
\mtmsbar\ using Eq.~(\ref{eq:fmt}) for the two calculations of
Fig.~\ref{fig:msbar_scenario1}.
The difference $\Delta m_{t}^{\overline{\rm MS}}$
between assuming that $\mtmc = \mtpole$ and
assuming $\mtmc\ = \mtmsbar$ is
given in Table~\ref{tab:msbar}.
We include
half of this difference symmetrically in the systematic
uncertainties and derive a value of $\mtmsbar = \msbarpa ^{+5.2}
_{-4.5}$~GeV using the calculation of Ref.~\cite{SMtheory_A} and
$\mtmsbar = \msbarpm ^{+5.1}_{-4.5}$~GeV using
Ref.~\cite{SMtheory_M}. 

To summarize, we extract the pole mass (Table~\ref{tab:pole}) and the
\msbar\ mass (Table~\ref{tab:msbar}) for
the top quark by comparing the measured
\sigmatt\ with different higher-order perturbative QCD
calculations. 
The Tevatron direct measurements of \mt\ are consistent with both
\mtpole\ measurements within 2~sd, but they are different by more than
2~sd from the extracted \mtmsbar.  
The results on \mtpole\ and their interplay with other
electroweak results within the SM are displayed in
Fig.~\ref{fig:w11}, which is based on Ref.~\cite{ew_fits}.

For the first time, \mtmsbar\ is extracted with the
\mt\ dependence of the measured \sigmatt\ taken into
account.
Our measurements favor the interpretation that the Tevatron \mt\
measurements based on reconstructing top quark decay products is
closer to the pole than to the \msbar\ top quark mass.

\section*{Acknowledgments}
We wish to thank M.~Cacciari, S.~Moch, M.~Neubert, M.~Seymour, and P.~Uwer 
for fruitful discussions regarding this analysis.
We would also like to thank M.~Cacciari, N.~Kidonakis, and L.~L.~Yang for 
providing us with their latest theoretical computations.

%
We thank the staffs at Fermilab and collaborating institutions,
and acknowledge support from the
DOE and NSF (USA);
CEA and CNRS/IN2P3 (France);
FASI, Rosatom and RFBR (Russia);
CNPq, FAPERJ, FAPESP and FUNDUNESP (Brazil);
DAE and DST (India);
Colciencias (Colombia);
CONACyT (Mexico);
KRF and KOSEF (Korea);
CONICET and UBACyT (Argentina);
FOM (The Netherlands);
STFC and the Royal Society (United Kingdom);
MSMT and GACR (Czech Republic);
CRC Program and NSERC (Canada);
BMBF and DFG (Germany);
SFI (Ireland);
The Swedish Research Council (Sweden);
and
CAS and CNSF (China).


\begin{thebibliography}{99}
\bibitem{jetmass_footnote}

An estimate for the mass parameter that appears in parton shower
algorithms can be obtained by speculating how an ideal all-order algorithm would
work ~\cite{mcpaper_formassdef} using the approach developed in
Ref.~\cite{jetmass_scheme}. Comparing parton shower results with
an all-order calculation, a relation between \mtpole\
and \mt\ can be derived. Hence, the pole mass
of the top quark could be about 1 GeV higher than $m_t$
from direct Tevatron measurements.
\end{thebibliography}
\end{document}